\documentclass[prd,nofootinbib,showkeys,showpacs]{revtex4}
\usepackage{graphicx}
\usepackage{amsmath}
\usepackage{dsfont}               

\newcommand{\sumint}{\hbox{$\sum$}\!\!\!\!\!\!\!\int }

\begin{document}

\title{Selfconsistent approximations, symmetries and choice of representation}
\author{Stefan Leupold}
\affiliation{Institut f\"ur Theoretische Physik, Universit\"at Giessen, Germany}

\begin{abstract}
In thermal field theory selfconsistent ($\Phi$-derivable) approximations are used
to improve (resum) propagators at the level of two-particle irreducible diagrams.
At the same time vertices are treated at the bare level. Therefore such approximations 
typically violate the Ward identities connected to internal symmetries. Examples are
presented how such violations can be tamed by a proper choice of representation for
the fields which describe the system under consideration. These examples cover the
issue of massless Goldstone bosons in the linear sigma model and the 
Nambu--Jona-Lasinio model and the problem of current conservation in theories with
massive vector mesons.
\end{abstract}

\maketitle

\section{Introduction and Summary}

For the description of quantum field theories in and also out of thermal equilibrium
$\Phi$-derivable approximations \cite{Luttinger:1960ua,Baym:1962sx,Cornwall:1974vz}
have gained a lot of attention in the last years \cite{Weinhold:1997ig,%
Ivanov:1998nv,vanHees:2000bp,vanHees:2001ik,vanHees:2001pf,vanHees:2002bv,Ivanov:2005bv,%
Blaizot:1999ap,Blaizot:2000fc,Blaizot:2001ev,Blaizot:2003br,Blaizot:2003an,%
Peshier:1998rz,Peshier:2000hx,Peshier:2001mk,%
Aarts:2002dj,Berges:2004pu,Alford:2004jj,Berges:2005hc,Berges:2004hn,%
Lenaghan:2000ey,Lenaghan:1999si,Roder:2003uz,Roder:2005vt}. Such an approach provides
a tool to go beyond purely perturbative calculations by resumming whole classes of 
diagrams. At the same time the approximation scheme is thermodynamically 
consistent \cite{Baym:1962sx,Ivanov:1998nv}. Typically the generating functional
which defines the approach is introduced as a functional of one- and two-point functions 
(classical fields and propagators). Consequently, the key quantity 
$\Phi$ is calculated from two-particle
irreducible (2PI) diagrams (see e.g.~\cite{Berges:2004pu} for a generalization
to $n$-PI diagrams). In this way, one deals with full propagators, while vertices
are treated at a perturbative level. If the system under consideration contains
internal symmetries, such an approach can violate the Ward identities connected to 
these symmetries \cite{vanHees:2002bv}. The reason is that Ward identities typically 
connect propagators and vertices. Therefore, problems can occur in a scheme in which
propagators are determined by involving diagrams of arbitrary loop order, while 
vertices are calculated only up to a given loop order. 

The purpose of the present work is to show some examples where such problems appear
and to point out how these problems can be avoided by a proper redefinition of fields.
The present work does {\em not} aim at a full formal solution of arbitrary problems 
connected
to resummations in the presence of internal symmetries. Rather for some examples of
practical relevance (Goldstone modes, current conservation in theories with 
massive vector mesons), it is shown how the problems can be tamed. Also the discussion
of renormalization issues (cf.~e.g.~\cite{vanHees:2001ik,vanHees:2001pf,vanHees:2002bv}) 
is beyond the scope of the present work.

In principle, physical quantities (e.g.~$S$-matrix elements) do not depend on 
the choice of representation \cite{Kamefuchi:1961sb,Fearing:1999fw}. 
Thus, it is important
to understand why a redefinition of fields, i.e.~a clever choice of representation
can help at all: The point is that 
typically one cannot calculate a physical quantity exactly, i.e.~one does
not have a full solution to the quantum field theoretical problem, but only an 
approximation. E.g.~in a perturbation theory differences originating from different
choices of representation are of higher order in the expansion parameter than the order 
studied (see e.g.~\cite{Scherer:1994wi} and references therein). 
In resummation schemes one typically involves classes
of processes/diagrams up to infinite order in the expansion parameter 
(e.g.~two-particle reducible diagrams) while other classes are treated perturbatively.
In such a scheme the only thing one can say about the (in-)dependence on the choice
of representation is that the results should become less dependent on the representation,
if one includes more and more processes/diagrams (in a systematic way). 
Therefore, in practice where one cannot solve the full quantum field theoretical problem,
the choice of representation might matter. As will be outlined in the present work 
it can indeed be used as a tool to improve the symmetry properties of a resummation
scheme. 

The rest of the present work is structured as follows: In the next section we
discuss the linear sigma model and in section \ref{sec:NJL} the Nambu--Jona-Lasinio
model. In both sections we focus on the problem that the propagator of the 
(supposed-to-be) Goldstone
bosons might not propagate massless modes, if it is calculated within a $\Phi$-derivable
approximation with resummed propagators but bare vertices. We will show that such a 
problem appears, if a linear representation for the Goldstone boson fields is used,
whereas Goldstone modes remain massless for a non-linear representation.
Note that a similar line of reasoning is also presented in \cite{Chanfray:2000ha},
however not in the context of $\Phi$-derivable schemes. For the Nambu--Jona-Lasinio
model also a redefinition of the quark fields will be important on top of the change
of representation for the Goldstone boson fields. In section \ref{sec:vm} we turn
to a different problem, namely the current (non-)conservation in theories with
massive vector mesons. Here it will turn out that the use of a tensor representation
for the vector mesons is superior to the frequently used vector representation. 
We will also present a projector formalism which deals with the technical aspects
of the tensor representation. 
As a side remark on the treatment of tadpoles in $\Phi$-derivable schemes we have added
an appendix.

\section{Linear sigma model}
\label{sec:lsm}

As a first example we take the $O(N+1)$ linear sigma model \cite{Goldstone:1962es}
\begin{equation}
  \label{eq:linsiglagr}
  {\cal L} = \frac12 \, \partial_\mu \vec \phi \, \partial^\mu \vec \phi
  + \frac12 \, m^2 \vec \phi\,^2
  - \frac{\lambda}{4} \, \left( \vec \phi\,^2 \right)^2 \,.
\end{equation}
Here $\vec \phi$ is a $N+1$ component vector (linear representation)
\begin{equation}
  \label{eq:linreplinsig}
  \vec \phi = \left(
    \begin{array}{c}
      \phi_0 \\ \vdots \\ \phi_N
    \end{array}
\right) \,.
\end{equation}
The Lagrangian (\ref{eq:linsiglagr}) is invariant with respect to the global
transformations
\begin{equation}
  \label{eq:trafosun}
  \vec \phi \to S \vec \phi
\end{equation}
with an arbitrary matrix $S \in O(N+1)$. 

For $m^2 > 0$ (and low enough temperatures) the system described by 
(\ref{eq:linsiglagr}) has a non-trivial ground 
state which spontaneously breaks the symmetry (\ref{eq:trafosun}). We choose the 
(positive) $\phi_0$ direction and find:
\begin{equation}
  \label{eq:symbreaklinsiglinrep}
  \phi_0^{\rm vac} = \sqrt{\frac{ m^2}{\lambda}} =: v  \,.
\end{equation}
To study the (quantum and thermal) fluctuations around this ground state we perform a
shift in (\ref{eq:linsiglagr})
\begin{equation}
  \label{eq:shiftlinsig}
  \phi_0 \to v + \phi_0
\end{equation}
which yields the Lagrangian
\begin{equation}
  \label{eq:linsiglinrep}
  {\cal L}_{\rm lin.\,repr.} = 
\frac12 \, \partial_\mu \vec \phi \, \partial^\mu \vec \phi
  - \frac{\lambda}{4} \, \left( \vec \phi\,^2 \right)^2
  - \frac12 \, m_0^2 \phi_0^2 
  - \lambda  v \, \phi_0 \, \vec \phi\,^2
\end{equation}
with the mass for the $\phi_0$ mode
\begin{equation}
  \label{eq:massphi0}
m_0^2 := 2 \lambda \,v^2 = 2 m^2  \,.
\end{equation}
Note that the spontaneous symmetry breaking induces a new three-point interaction 
term, the last term on the right hand side of (\ref{eq:linsiglinrep}).

\begin{figure}[htbp]
    \includegraphics[keepaspectratio,width=0.8\textwidth]{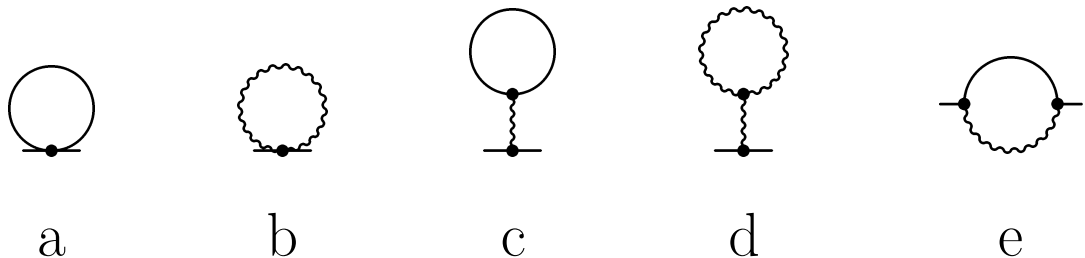}
  \caption{Self energy contributions which --- in a perturbative treatment ---
conspire such that no mass terms are induced. Solid lines denote $\phi_i$ modes
($i \neq 0$), wiggly lines $\phi_0$ modes. See main text for details.}
  \label{fig:selfphi4}
\end{figure}

Spontaneously broken global symmetries cause the appearance of massless Goldstone 
modes \cite{Goldstone:1961eq,Goldstone:1962es}. 
For the studied system these are the $\phi_i$ ($i\neq 0$) modes. 
Indeed,
on the tree level there are no mass terms for the $\phi_i$ modes. Single
loop diagrams induce mass terms. In a perturbative expansion, however, such mass
terms cancel in the sum of all contributing loop diagrams. An example of such a
cancellation is depicted in figure \ref{fig:selfphi4}.
%
There, all one-loop self energy diagrams for the $\phi_i$ modes are shown. 
The sum of these contributions is proportional to $I_a + I_b + I_c + I_d + I_e$
with the $\phi_i$ snail diagram
\begin{equation}
  \label{eq:isnail}
  I_a = \lambda \, (N + 2) \, \sumint \frac{d^{4}p }{ (2 \pi)^{4}} \,
          \frac{1}{p^2}  \,,
\end{equation}
the $\phi_0$ snail diagram
\begin{equation}
  \label{eq:0snail}
  I_b = \lambda \, \sumint  \frac{d^{4}p }{ (2 \pi)^{4}} \,
          \frac{1}{p^2-m_0^2}  \,,
\end{equation}
the $\phi_i$ tadpole
\begin{equation}
  \label{eq:itadp}
  I_c = (\lambda v)^2 \, 2 N \, \frac{1}{-m_0^2} \,
     \sumint  \frac{d^{4}p }{ (2 \pi)^{4}} \, \frac{1}{p^2}
= - \lambda N \, \sumint  \frac{d^{4}p }{ (2 \pi)^{4}} \, \frac{1}{p^2} \,,
\end{equation}
the $\phi_0$ tadpole
\begin{equation}
  \label{eq:0tadp}
 I_d = 6 \, (\lambda v)^2 \, \frac{1}{-m_0^2} \, \sumint  \frac{d^{4}p }{ (2 \pi)^{4}} \,
          \frac{1}{p^2-m_0^2} 
= - 3 \lambda \, \sumint  \frac{d^{4}p }{ (2 \pi)^{4}} \, \frac{1}{p^2-m_0^2}
\end{equation}
and the $\phi_i$-$\phi_0$ loop 
\begin{equation}
  \label{eq:loop}
  I_e(k) = 4 \, (\lambda v)^2 \, \sumint  \frac{d^{4}p }{ (2 \pi)^{4}} \, 
   \frac{1}{(p^2-m_0^2) \, (p-k)^2} 
= 2 \lambda m_0^2 \, \sumint  \frac{d^{4}p }{ (2 \pi)^{4}} \, 
   \frac{1}{(p^2-m_0^2) \, (p-k)^2}  \,.
\end{equation}
We have introduced the Matsubara formalism \cite{Matsubara:1955ws}
to calculate the diagrams at finite temperature. In the following, our considerations
will be at a purely formal level. Therefore we are not concerned with the renormalization
of the expressions in (\ref{eq:isnail})-(\ref{eq:loop}).

A mode remains massless, if the corresponding self energy $\Pi$ satisfies
\begin{equation}
  \label{eq:condmassless}
\lim\limits_{k \to 0} \Pi(k) = 0 \,.
\end{equation}
The following decomposition for diagram (e) ensures that the $\phi_i$ modes remain 
massless:
\begin{eqnarray}
  \label{eq:decompfocksnail}
\frac{m_0^2}{(p^2-m_0^2) \, (p-k)^2} 
= \frac{1}{p^2-m_0^2} - \frac{1}{(p-k)^2} - \frac{-2kp + k^2}{(p^2-m_0^2) \, (p-k)^2}
\end{eqnarray}
i.e.~the first two terms on the right hand side of \eqref{eq:decompfocksnail}
cancel the result of the sum of $I_a + I_b + I_c + I_d$ whereas the last term 
vanishes with the external momentum $k$. 

\begin{figure}[htbp]
    \includegraphics[keepaspectratio,width=0.5\textwidth]{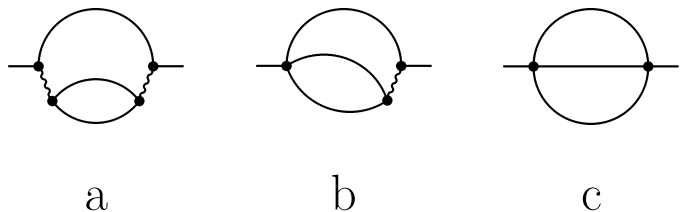}
  \caption{Some two-loop self energy contributions for the linear sigma model. 
Solid lines denote $\phi_i$ modes ($i \neq 0$), wiggly lines $\phi_0$ modes.}
  \label{fig:selfphi4-two}
\end{figure}
With the decomposition (\ref{eq:decompfocksnail}) one basically transforms
two propagators to one propagator, i.e.~one contracts one propagator in diagram
\ref{fig:selfphi4}(e). The remaining piece vanishes with the external momentum,
i.e.~does not play a role. By the contraction of one propagator
diagram \ref{fig:selfphi4}(e) looks the same as diagram \ref{fig:selfphi4}(a) or
(b), respectively, depending which propagator is contracted. Also at higher loop order
corresponding cancellations happen 
between diagrams which differ from each other by one propagator. As an illustrative
example suppose that one wants
to consider the two-loop diagram shown in figure \ref{fig:selfphi4-two}(a). To
ensure that the mode remains massless one needs in addition (besides others) the 
diagrams depicted in figure \ref{fig:selfphi4-two}(b) and (c). 

\begin{figure}[htbp]
    \includegraphics[keepaspectratio,width=0.8\textwidth]{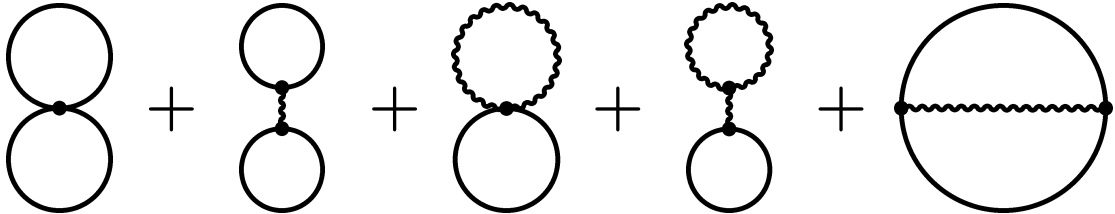}
  \caption{Contributions to the $\Phi$-functional (relevant for the generation of
$\phi_i$ self energies) in two-loop order for the 
linear sigma model (in linear
representation). Solid lines denote $\phi_i$ modes
($i \neq 0$), wiggly lines $\phi_0$ modes. See main text for details.}
  \label{fig:Phiphi4}
\end{figure}

This rather subtle cancellation does not work any longer
once the propagators are dressed. In the following, we will demonstrate the problem in 
two ways. Suppose that we calculate the $\phi_i$ propagators 
according
to a $\Phi$-derivable approximation using for $\Phi$ the two-loop diagrams depicted
in figure \ref{fig:Phiphi4}.\footnote{Note that figure \ref{fig:Phiphi4} shows
only the contributions to $\Phi$ which generate self energies for the $\phi_i$ modes.
In other words, diagrams which consist solely of $\phi_0$ propagators are not 
displayed explicitly.} 
This approach generates all self energies 
shown in figure \ref{fig:selfphi4} with the important difference that now all internal 
lines should be regarded as full instead of bare propagators.\footnote{Actually the
vertical wiggly line in the tadpole diagrams is still a bare propagator. This
subtlety is discussed in the appendix.} The necessary cancellation between  diagram
(e) and the sum of the others would still take place, {\em if} the following difference
vanished with the external momentum $k$ (cf.~equation (\ref{eq:decompfocksnail})):
\begin{eqnarray}
  \label{eq:decompfocksnail-full}
\Delta := \frac{m_0^2}{[p^2-m_0^2-\Pi_0(p)] \, [(p-k)^2-\Pi_i(p-k)]} 
- \left( \frac{1}{p^2-m_0^2-\Pi_0(p)} - \frac{1}{(p-k)^2-\Pi_i(p-k)} \right) \,,
\end{eqnarray}
where we have introduced self energies $\Pi_0$ and $\Pi_i$ for the $\phi_0$ and
$\phi_i$ modes, respectively. Instead of an expression which vanishes with $k$
we get
\begin{equation}
  \label{eq:Deltanonvan}
\Delta 
= \frac{2 kp - k^2 + \Pi_i(p-k)-\Pi_0(p)}{[p^2-m_0^2-\Pi_0(p)] \, [(p-k)^2-\Pi_i(p-k)]} 
\stackrel{k \to 0}{\to} 
\frac{\Pi_i(p)-\Pi_0(p)}{[p^2-m_0^2-\Pi_0(p)] \, [p^2-\Pi_i(p)]} \,.
\end{equation}
In general, the right hand side of (\ref{eq:Deltanonvan}) does not vanish, since
the self energies $\Pi_0$ and $\Pi_i$ are not the same. E.g.~the self energy for
the $\phi_0$ mode has an imaginary part coming from the decay into two (ideally massless)
$\phi_i$ modes. This decay channel is not present for a $\phi_i$ mode.

There is a second way to see that the cancellation does not work any more. For that 
purpose we work out which {\em perturbative} diagrams are generated from the $\Phi$
functional of figure \ref{fig:Phiphi4} and which are not. Obviously, one
generates (besides infinitely many 
other perturbative diagrams) the (perturbative!) two-loop 
self energy shown
in figure \ref{fig:selfphi4-two}(a). (Note that the lines in figure \ref{fig:Phiphi4}
denote {\em full} propagators, whereas the lines in figure \ref{fig:selfphi4-two}
denote {\em bare} propagators.) On the other hand, the diagrams depicted in
figure \ref{fig:selfphi4-two}(b) and (c) are {\em not} generated from $\Phi$ as given
in figure \ref{fig:Phiphi4}. Three-loop diagrams for $\Phi$ would be necessary here.
As already pointed out, all diagrams of figure \ref{fig:selfphi4-two} would be needed
to ensure that the $\phi_i$ modes remain massless at the two-loop level.

More generally, the symmetries (which dictate the appearance
of Goldstone modes) lead to Ward identities which typically connect propagators and 
vertices. If one resums propagators to all orders but truncates the vertices, one
might get problems, since the necessary cancellation of different diagrams is not
ensured any longer. Note that e.g.~diagram \ref{fig:selfphi4-two}(a) is a propagator
correction to diagram \ref{fig:selfphi4}(e), while diagram \ref{fig:selfphi4-two}(b)
is a vertex correction. Also note that the inclusion of diagram \ref{fig:selfphi4-two}(b)
--- even with full propagators --- would not solve the problem in a 2PI $\Phi$-derivable
scheme: Using full propagators one would need a {\em full} vertex and not just bare
plus one-loop. 

The previous discussion shows that one has to look for a formalism where such 
cancellations are not needed.
To see that this is indeed possible and practically conceivable we turn
to the non-linear representation (see e.g.~also \cite{Chanfray:2000ha})
\begin{equation}
  \label{eq:nonlin-linsig}
\vec \phi = \sigma \vec U
\end{equation}
with $\vec U\,^2 = 1$. Obviously, as a unit vector in $N+1$ dimensions, $\vec U$ 
contains $N$ degrees of freedom which we will call ``$\pi$ modes'' in the following.
In the non-linear representation the
Lagrangian (\ref{eq:linsiglagr}) takes the form
\begin{equation}
  \label{eq:linsignonlinlagr}
{\cal L} = \frac{1}{2} \, \sigma^2 \, \partial_\mu \vec U \, \partial^\mu \vec U
+ \frac12 \, \partial_\mu \sigma \partial^\mu \sigma
+ \frac12 \, m^2 \, \sigma^2 - \frac{\lambda}{4} \, \sigma^4  \,.
\end{equation}
Spontaneous symmetry breaking induces a finite vacuum expectation value for the $\sigma$
mode:
\begin{equation}
  \label{eq:sigexpval}
  \sigma_{\rm vac} = \sqrt{\frac{ m^2}{\lambda}} = v
\end{equation}
which of course agrees with (\ref{eq:symbreaklinsiglinrep}). With the shift
\begin{equation}
  \label{eq:shiftnonlin}
  \sigma \to v + \sigma
\end{equation}
one gets the Lagrangian
\begin{eqnarray}
  \label{eq:linsignonlinrep}
  {\cal L}_{\rm non-lin.\,repr.} & = &
\frac{1}{2} \, v^2 \, \partial_\mu \vec U \, \partial^\mu \vec U
+ v \, \sigma \, \partial_\mu \vec U \, \partial^\mu \vec U
+\frac{1}{2} \, \sigma^2 \, \partial_\mu \vec U \, \partial^\mu \vec U
+ \frac12 \, \partial_\mu \sigma \partial^\mu \sigma
- m^2 \sigma^2 - \lambda v \sigma^3  
- \frac{\lambda}{4} \, \sigma^4  \,.
\end{eqnarray}

Obviously, the $\pi$ modes appear only with derivatives. Therefore, all interactions
vanish in the limit of soft momenta and no mass terms are induced. This statement
is separately true for each conceivable Feynman diagram for the $\pi$ self energy. 
Therefore no subtle cancellations are needed to ensure the appearance of Goldstone
modes. In any $\Phi$-derivable scheme based on (\ref{eq:linsignonlinrep})
the $\pi$ modes remain massless. Therefore a non-linear representation is better
suited for such a resummation scheme. For practical applications the unit vector
$\vec U$ must be expanded in powers of the pion field. This is completely analogous
to the non-linear sigma model and its extention to chiral perturbation 
theory \cite{gasleut1}. We will not elaborate on this issue here any more.
We note, however, that in spite of the chosen non-linear representation in
(\ref{eq:linsignonlinrep}) this Lagrangian still describes the linear sigma model
since the sigma mode is not frozen, but has a tree level mass as 
given in (\ref{eq:massphi0}).

\section{Nambu--Jona-Lasinio model}
\label{sec:NJL}

As a second example we study the Nambu--Jona-Lasinio (NJL) 
model \cite{Vogl:1991qt,Klevansky:1992qe,Hatsuda:1994pi}. It is widely used
as a quark model which possesses the chiral symmetry of QCD. The spontaneous breakdown
of this symmetry in vacuum and its restoration at finite temperatures and/or
baryon densities has been studied extensively within the NJL model. Also the appearance
of Goldstone modes can be studied explicitly within this model. For simplicity we
restrict ourselves in the following to two quark flavors. One way to write down the
Lagrangian is \cite{Klevansky:1992qe}
\begin{equation}
  \label{eq:NJLbos}
  {\cal L} = 
\bar q \, \left( i \gamma_\mu \partial^\mu -m 
- \sigma - i \gamma_5 \pi^a \tau^a \right) q
- \frac{1}{4G} \, \left(\sigma^2 + \vec \pi^2 \right)
\end{equation}
with the current quark mass $m$.\footnote{We assume 
perfect isospin symmetry for simplicity.} Integrating out the sigma and pion fields
(which posses no dynamics at tree level) one obtains the usual NJL Lagrangian with its
four-quark couplings. 

\begin{figure}[htbp]
    \includegraphics[keepaspectratio,width=0.07\textwidth]{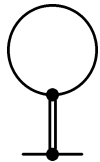}
  \caption{Hartree contribution to the quark self energy. The solid lines denote quarks.
    The double line
    denotes a bare $\sigma$ propagator as obtained from
    the Lagrangian (\ref{eq:NJLbos}), i.e.~it is just $2G$.}
  \label{fig:selfNJLquark}
\end{figure}

\begin{figure}[htbp]
    \includegraphics[keepaspectratio,width=0.5\textwidth]{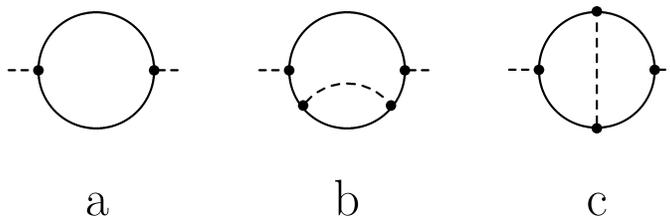}
  \caption{Perturbative self energy contributions for the pion. Solid lines denote 
    quarks, dashed lines pions.}
  \label{fig:selfNJLpion}
\end{figure}

The NJL model possesses
a systematic expansion scheme, namely an expansion in inverse powers of the number of
quark colors $N_c$. In that context we note that
$G \sim 1/N_c$ (see e.g.~\cite{Klevansky:1992qe}). In leading order of the $1/N_c$ 
expansion the quarks get a dynamically generated constituent mass 
(Hartree approximation). 
The corresponding quark self energy is depicted in figure \ref{fig:selfNJLquark}.
Note that the solid line in the loop is supposed to be a full quark propagator, 
i.e.~''bubbles within bubbles'' are implicitly generated by this diagram.
For the mesons, e.g.~the pion, the leading order contribution to the self energy is
given by the one-loop diagram shown in figure \ref{fig:selfNJLpion}(a). Again the
solid lines denote full (here Hartree) quark propagators.
Note that this quark loop yields also a kinetic term for the pion 
which is obviously not present at tree level, i.e.~in the Lagrangian (\ref{eq:NJLbos}). 
This kinetic term is proportional
to the square of the pion decay constant (see e.g.~\cite{Klevansky:1992qe}). 

\begin{figure}[htbp]
    \includegraphics[keepaspectratio,width=0.5\textwidth]{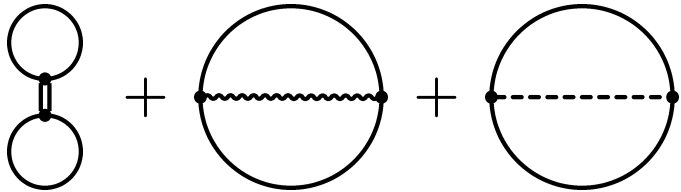}
  \caption{$\Phi$-functional up to $O(N_c^0)$ for the linear 
    representation (\ref{eq:NJLbos}). All lines denote full propagators except for
    the double line which is a bare sigma propagator. (This subtlety is explained in
    the appendix.) Solid lines denote quarks, 
    dashed lines pions
    and wiggly lines sigmas.}
  \label{fig:Philin}
\end{figure}

At leading $1/N_c$ order the dynamics of the quarks (the generation of a constituent
quark mass) influences the meson properties. On the other hand, the dynamics of the 
mesons are not fed back to influence the quark properties. Actually it is only
the expectation value of the sigma, i.e.~the one-point function, which causes the
Hartree diagram. The connection of tadpoles and one-point functions is discussed from
a somewhat more general point of view in the appendix. It is only at next-to-leading
order where the meson propagators influence the quark properties. Therefore,
processes like quark-quark or quark-meson scattering come into play only at
next-to-leading order of the $1/N_c$ expansion. On the other hand, such processes
are of interest e.g.~for a dynamical description of the chiral phase transition
\cite{Rehberg:1998me}. From the point of view of a $\Phi$-derivable approximation this
looks as follows: At leading order in $1/N_c$ one only has the first (left) diagram
shown in figure \ref{fig:Philin}. It is $O(N_c)$ and generates the Hartree self energy
of figure \ref{fig:selfNJLquark}. 
The next-to-leading order contributions $O(N_c^0)$ come from the other two diagrams
of figure \ref{fig:Philin}. To generate the mesons and couple the dynamics of quarks 
and mesons in a selfconsistent way one has to involve (at least) all the diagrams of 
figure \ref{fig:Philin}.
However, as we will discuss next, such a selfconsistent scheme again has problems to 
ensure that the pions keep their character as Goldstone bosons.

Obviously there is a mass term $1/(2G)$ for the pion at tree level. In the chiral
limit ($m = 0$) this mass term is canceled exactly
by the quark loop depicted in figure \ref{fig:selfNJLpion}(a),
{\em if} the quark propagator is determined at the Hartree level 
depicted in figure \ref{fig:selfNJLquark}.
For finite quark masses one obtains the Gell-Mann--Oakes--Renner 
relation \cite{GOR}.
The cancellation between tree level
and quark loop does not work any more, if the quark propagator is changed beyond
the Hartree level without changing the corresponding vertices. E.g.~the $\Phi$-derivable
approximation depicted in figure \ref{fig:Philin} generates the perturbative 
contribution shown in 
figure \ref{fig:selfNJLpion}(b)
but {\em not} the diagram of figure \ref{fig:selfNJLpion}(c). 
Only both diagrams (b) and (c) of figure \ref{fig:selfNJLpion}
ensure that the pion remains massless (in the chiral limit). In the following, we
will demonstrate how this problem can be circumvented.

Now we turn to a non-linear realization by identifying
\begin{equation}
  \label{eq:linnonlinNJL1}
\sigma + i \tau^a \pi^a = \tilde \sigma e^{i \tau^a \tilde \pi^a/F} =: 
\tilde \sigma U \,.
\end{equation}
We have introduced the pion decay constant $F$. To be more specific, at present 
$F$ is an
arbitrary parameter which drops out of all physical quantities. When a kinetic
term for the pion is generated from the loops the free parameter is properly replaced
by the pion decay constant $F_\pi$. 
Inserting (\ref{eq:linnonlinNJL1}) in (\ref{eq:NJLbos}) we get
\begin{eqnarray}
  \label{eq:NJLbos2}
  {\cal L} & = &
\bar q  \left( i \gamma_\mu \partial^\mu -m 
- \tilde \sigma U P_R - \tilde \sigma U^\dagger P_L \right) q
- \frac{1}{4G} \, \tilde \sigma^2 
\nonumber \\ 
& = &
 \bar{q_R} i \gamma_\mu \partial^\mu q_R + \bar{q_L} i \gamma_\mu \partial^\mu q_L 
- \bar{q_L}  \left( \tilde\sigma U + m \right) q_R
- \bar{q_R}  \left( \tilde\sigma U^\dagger + m \right) q_L
- \frac{1}{4G} \, \tilde \sigma^2 
\end{eqnarray}
where we have introduced chiral projectors $P_{R/L} := \frac12 (1 \pm \gamma_5)$
and right/left handed quarks $q_{R/L} := P_{R/L} q$. Obviously, in this non-linear
representation of the boson fields there is no pion mass at the tree level. Still
it might happen that a pion mass is generated by loops. In the following, we will
demonstrate how to avoid that. To this end, we also change
the representation for the quark fields \cite{Manohar:1983md,espraf}:
\begin{equation}
  \label{eq:quarkconst}
q_R' = U^{1/2} q_R \, \qquad q_L' = U^{-1/2} q_L \,.
\end{equation}
This yields
\begin{equation}
  \label{eq:NJLbos3}
  {\cal L}  = 
\bar q' \, i \gamma_\mu \partial^\mu  q' 
+ \bar{q'_R} \, i U^{1/2} \gamma_\mu (\partial^\mu U^{-1/2}) q'_R
+ \bar{q'_L} \, i U^{-1/2} \gamma_\mu (\partial^\mu U^{1/2}) q'_L
- \bar{q'_L}  \left( \tilde\sigma  + m U^\dagger \right) q'_R
- \bar{q'_R}  \left( \tilde\sigma  + m U \right) q'_L
- \frac{1}{4G} \, \tilde \sigma^2 \,.
\end{equation}
In the following, we will only be concerned with
the non-linear representation. Therefore we drop from now on the tilde and prime
assignment to the fields in (\ref{eq:linnonlinNJL1}) and (\ref{eq:quarkconst}). 
Obviously, in (\ref{eq:NJLbos3}) all interactions between the pion fields 
(encoded in $U$) and the quarks come with derivatives of the pion fields
or with the current quark mass. Thus, in the chiral limit soft pions decouple
from the quarks. No mass terms for the pion can therefore be generated in any loop
order (in the chiral limit). No cancellation of diagrams is needed to achieve that
property, since each interaction vertex separately ensures the decoupling of pions.

\begin{figure}[htbp]
    \includegraphics[keepaspectratio,width=0.7\textwidth]{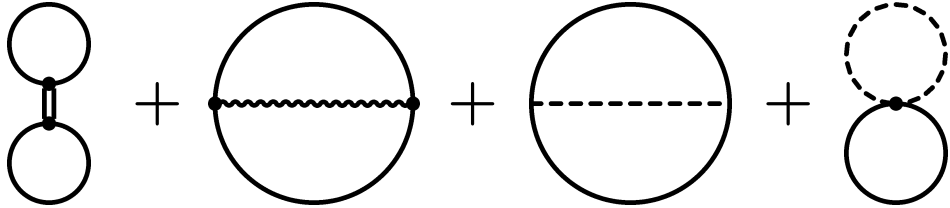}
  \caption{$\Phi$-functional up to $O(N_c^0)$ for the non-linear 
    representation (\ref{eq:NJLbos4}). All lines denote full propagators except for
    the double line which is a bare sigma propagator (cf.~figure \ref{fig:Philin} and
    the appendix). Solid lines denote quarks, 
    dashed lines pions
    and wiggly lines sigmas. The three-point pion-quark vertex
    consists of two parts depicted in figure \ref{fig:pionvertices}.}
  \label{fig:Phinonlin}
\end{figure}

In powers of $1/N_c$ the $\Phi$-derivable approximation shown in 
figure \ref{fig:Philin} (using the linear
representation (\ref{eq:NJLbos})) yields the generating functional up to $O(N_c^0)$.
The same accuracy can be obtained in the non-linear representation, if all $U$'s
appearing in (\ref{eq:NJLbos3}) are expanded up to $O(1/F^2)$. Note that 
$F^2 \sim N_c$ which can be easily obtained from the Gell-Mann--Oakes--Renner relation.
We obtain
\begin{eqnarray}
  \label{eq:NJLbos4}
  {\cal L} & \approx &
\bar q  \left( i \gamma_\mu \partial^\mu - m - \sigma 
+ \frac{1}{2F} \, \gamma_\mu \partial^\mu \pi^a \tau^a \gamma_5 
+ \frac{m}{2F^2} \, \vec \pi^2 + \frac{i m}{F} \, \gamma_5 \pi^a \tau^a
\right) q
- \frac{1}{4G} \,  \sigma^2 \,.
\end{eqnarray}
Actually there emerges one more term with two pion fields, the Weinberg-Tomozawa
term $\sim \bar q \gamma^\mu \epsilon^{abc} \pi ^a \partial_\mu \pi^b \tau^c q$.
On the level of approximation which we treat in figure \ref{fig:Phinonlin}, 
this term enters the calculation 
of $\Phi$ only as a tadpole type contribution. Since
the Weinberg-Tomozawa term is flavor changing, this tadpole vanishes (as long as there 
is no isospin chemical potential). 
Of course, this would change, if we were interested in a calculation of 
$\Phi$ beyond $O(N_c^0)$. 
\begin{figure}[htbp]
    \includegraphics[keepaspectratio,width=0.5\textwidth]{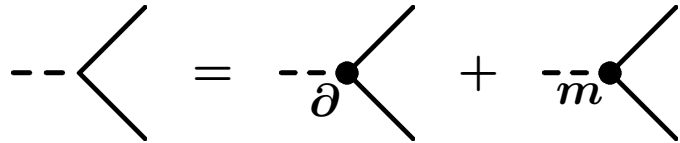}
  \caption{Pion-quark vertices of the non-linear 
    representation (\ref{eq:NJLbos4}). Solid lines denote quarks, dashed lines pions.}
  \label{fig:pionvertices}
\end{figure}
\begin{figure}[htbp]
    \includegraphics[keepaspectratio,width=0.7\textwidth]{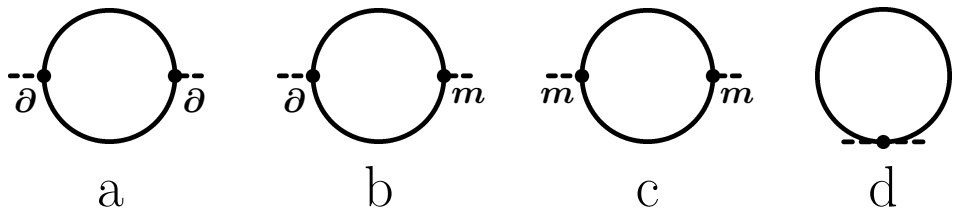}
  \caption{Pion self energy obtained in the non-linear representation from the
    $\Phi$-functional shown in figure \ref{fig:Phinonlin}. 
    Note that all internal lines denote full 
    propagators.}
  \label{fig:selfNJLpionnl}
\end{figure}

The pion mass $M_\pi$ is obtained from the self energy diagrams depicted in 
figure \ref{fig:selfNJLpionnl}. Obviously, diagrams (a), (b), (c) and (d) are 
proportional to $M_\pi^2$, $m M_\pi$, $m^2$ and $m$, respectively. Thus, comparing
diagrams (a) and (d) yields $M_\pi^2 \sim m$ and diagrams (b) and (c) are only
subleading corrections. Indeed, it is easy to see that the correct 
Gell-Mann--Oakes-Renner relation emerges from diagrams (a) and (d): The
quark condensate appears in diagram (d), together with the vertex $\sim m$. 
On the other hand, diagram (a) is caused by the pseudovector
interaction in (\ref{eq:NJLbos4}) which couples the derivative of the pion field
to the axial-vector current. The latter defines the pion decay constant. Thus,
$M_\pi^2 F_\pi^2$ emerges from diagram (a). We conclude that the Goldstone boson
character of the pion is respected in a $\Phi$-derivable approach to the NJL model, 
if the starting point is the Lagrangian (\ref{eq:NJLbos3}) instead of
(\ref{eq:NJLbos}). Therefore, the non-linear representation is clearly better suited
for studies of the low-temperature phase where chiral symmetry is spontaneously broken.
However, it is important to note that there is one aspect where
the linear representation has its merits: In (\ref{eq:NJLbos}) one can clearly see the 
chiral partners of the model, $\sigma$ and $\vec \pi$, which become degenerate for
temperatures above the phase transition \cite{Klevansky:1992qe}. Naturally, a 
non-linear representation does not display the chiral partners so explicitly. 
What makes the use of the non-linear representation at the technical level rather 
involved close to or even above the phase transition can be understood as follows:
We have argued that a proper expansion parameter is $1/F$ to come from (\ref{eq:NJLbos3})
to the practically useful Lagrangian (\ref{eq:NJLbos4}). Close to the chiral transition
the relevant $F$ becomes small, however. Therefore, an expansion in $1/F$ ceases to be
useful. A numerical study of the $\Phi$-functional shown in figure \ref{fig:Phinonlin}
as a function of the temperature is beyond the scope of the present work. 
At least for temperatures below the chiral transition the non-linear representation
is an appropriate tool as it 
respects the Goldstone boson character of the pions also within resummation schemes.

\section{Vector mesons and current conservation}
\label{sec:vm}

We now turn to a different problem encountered in resummation schemes, namely the 
violation of 
current conservation for systems with massive vector mesons. Typically vector mesons
are described by a Lorentz vector field $V_\mu$ which has four indices. On the other hand,
a massive vector meson has only three polarizations. The (free) equation of motion 
(Proca equation \cite{Mosel:1989jf}) is 
then constructed such that one component of the vector field is frozen by the condition
\begin{equation}
  \label{eq:nolong}
  \partial^\mu V_\mu  = 0 \,.
\end{equation}
If interactions are switched on and the
vector mesons are only coupled to currents $j_\mu$ which are conserved, 
then \eqref{eq:nolong}
remains valid in the presence of these interactions --- at least for a full solution
of the quantum field theoretical problem. However, the conservation of $j_\mu$ is
typically a consequence of an internal symmetry which on the level of $n$-point functions
connects propagators and vertices (an example is discussed in \cite{vanHees:2000bp}). 
For approximate solutions and especially
using resummation schemes it might therefore appear that current conservation is spoiled.
As a consequence also \eqref{eq:nolong} is violated and the current non-conservation
is proliferated by the now non-vanishing longitudinal mode of the vector meson.
Recipes how to tame such problems and the appearance of possible new problems caused by 
the use of such recipes are 
discussed in \cite{vanHees:2000bp,Riek:2004kx,Ruppert:2004yg,Riek:2006vq,Ruppert:2006he}.
In the following we will demonstrate how the problem of current non-conservation
can be circumvented by a different choice of representation for the vector meson field.

We start with the vector meson Lagrangian
\begin{equation}
  \label{eq:lagrvec}
{\cal L} = - \frac14 \, F_{\mu\nu} F^{\mu\nu} + \frac12 \, m^2 V_\mu V^\mu + 
V_\mu j^\mu 
\end{equation}
with a source term $j^\mu$ and the field strength 
$F_{\mu\nu} = \partial_\mu V_\nu - \partial_\nu V_\mu$. Interactions of the vector
mesons with other fields can be encoded in $j^\mu$. Note that typical interactions
of vector mesons with other fields can be written in the 
way (\ref{eq:lagrvec}). However, it is not the most general case: Also terms like 
$V_\mu V_\nu j^{\mu\nu}$ are conceivable. Indeed, if vector meson masses are
generated via a Higgs field $\phi$ \cite{pesschr}, terms 
with $j^{\mu\nu} \sim g^{\mu\nu} \phi^2$ appear. Terms with two vector fields
in the interaction part are unfortunately not suitable for the field redefinitions 
which we discuss below. Therefore, our framework does not cover the case of dynamical
vector meson mass generation and, as we will see below, also not the case of
massless vector mesons. Nonetheless, the Lagrangian (\ref{eq:lagrvec}) covers a large
body of frequently used hadronic Lagrangians with vector mesons.
Concerning field redefinitions
for vector states we also refer to appendix B in \cite{Borasoy:1995ds}.

We introduce a new antisymmetric tensor field $\bar F_{\mu\nu}$ and study the new
Lagrangians
\begin{equation}
  \label{eq:lagrvec2}
{\cal L}' =  \frac14 \, \bar F_{\mu\nu} \bar F^{\mu\nu} + \frac12 \, m^2 V_\mu V^\mu 
- \bar F_{\mu\nu} \partial^\mu V^\nu  + V_\mu j^\mu
\end{equation}
and
\begin{equation}
  \label{eq:lagrvec3}
{\cal L}'' =  \frac14 \, \bar F_{\mu\nu} \bar F^{\mu\nu} + \frac12 \, m^2 V_\mu V^\mu 
+ V^\mu \left(j_\mu + \partial^\nu \bar F_{\nu\mu} \right)  \,.
\end{equation}
Since $\bar F_{\mu\nu}$ appears at most in quadratic order in (\ref{eq:lagrvec2})
this field can be integrated out. Since there are no derivatives acting on this field
one obtains again a local action. Actually, one ends up with the 
original Lagrangian (\ref{eq:lagrvec}). On the other hand, the Lagrangians
(\ref{eq:lagrvec2}) and (\ref{eq:lagrvec3}) differ only by an irrelevant total
derivative. Thus all three Lagrangians are equivalent.
In (\ref{eq:lagrvec3}) the field $V_\mu$ appears only up to quadratic order and
without derivatives acting on it. Hence, $V_\mu$ can be easily integrated out.
One gets
\begin{equation}
  \label{eq:lagrvec4}
{\cal L}'' \to  \frac14 \, \bar F_{\mu\nu} \bar F^{\mu\nu} 
- \frac{1}{2 m^2} \, \partial^\nu \bar F_{\nu\mu} \partial_\alpha \bar F^{\alpha \mu}
- \frac{1}{m^2} j^\mu \partial^\nu \bar F_{\nu\mu}
- \frac{1}{2 m^2} \, j_\mu j^\mu  \,.
\end{equation}
To achieve a proper normalization of the kinetic terms we 
introduce $\bar F_{\mu\nu} =: m V_{\mu\nu}$ and obtain
\begin{equation}
  \label{eq:lagrvec5}
{\cal L}_{\rm tensor\,repr.} =  \frac14 \, m^2 V_{\mu\nu} V^{\mu\nu} 
- \frac{1}{2} \, \partial^\nu V_{\nu\mu} \partial_\alpha V^{\alpha \mu}
- \frac{1}{m} j^\mu \partial^\nu V_{\nu\mu}
- \frac{1}{2 m^2} \, j_\mu j^\mu  \,.
\end{equation}
The first two terms on the right hand side are just the mass and kinetic term of
the free Lagrangian for vector fields in the tensor 
representation \cite{gasleut1,eckgas,Ecker:1989yg}. The other two terms represent
the interaction of the vector field with the source and a quadratic source term.
The latter induces a point interaction, if the source is expressed in terms of the
fields the vectors are supposed to interact with. 

In contrast to the original interaction term $V_\mu j^\mu$ the new interaction
term $j^\mu \partial^\nu V_{\nu\mu}$ automatically projects on transverse states:
\begin{equation}
  \label{eq:projtrans}
j^\mu \partial^\nu V_{\nu\mu} 
= j^\mu \bigg( 
     \underbrace{g_{\mu\alpha} 
                 -\frac{\partial_\mu \partial_\alpha}{\partial^2}}_{P^T_{\mu\alpha}} 
      +\underbrace{\frac{\partial_\mu \partial_\alpha}{\partial^2}}_{P^L_{\mu\alpha}} 
        \bigg)
  \partial_\nu V^{\nu\alpha}
= j^\mu P^T_{\mu\alpha} \partial_\nu V^{\nu\alpha}
\end{equation}
where we have used
\begin{equation}
  \label{eq:antisymdrop}
\partial_\alpha \partial_\nu V^{\nu \alpha} = 0
\end{equation}
which holds since $V^{\nu \alpha}$ has been introduced as an antisymmetric tensor.
Obviously problems with current conservation are now no longer proliferated by the
vector mesons. Therefore a tensor representation for the vector mesons provides a
better starting point for selfconsistent approximations than the frequently used
vector representation.

Trading a vector field with one Lorentz index with a tensor field with two indices, one
might get the feeling that in practice this becomes technically rather difficult. 
However, such kind of difficulties always have a simple solution: One just needs
the proper projectors to decompose everything in scalar quantities (times projectors).
Without much details we present in the following the projectors for the tensor
representation which are required for (equilibrium) in-medium calculations, i.e.~for
a situation where one has a Lorentz vector $p$ which specifies the medium. In such a
case one can distinguish vector mesons which move with respect to the medium from
those which do not. (In vacuum one can always boost to the frame where the vector
meson is at rest.) For moving (massive) vector mesons one can distinguish their 
respective polarization \cite{Gale:1990pn}: 
Either it is longitudinal (l) or transverse (t) with respect to the
{\em three}-momentum of the vector meson. (We recall that the polarization is always
transverse (T) with respect to the {\em four}-momentum.) 

The pertinent tensor structures are: the unity tensor
\begin{equation}
  \label{eq:proj1}
  P_1^{\mu \nu \alpha \beta} = 
  \frac12 \left( g^{\mu \alpha} g^{\nu \beta} - g^{\mu \beta} g^{\nu \alpha} \right) \,,
\end{equation}
the tensor transverse with respect to the four-momentum $k$
\begin{equation}
  \label{eq:projT}
  P_T^{\mu \nu \alpha \beta} = \frac{1}{2k^2} 
  \left( g^{\mu \alpha} k^\nu k^\beta - g^{\mu \beta} k^\nu k^\alpha 
    - g^{\nu \alpha} k^\mu k^\beta + g^{\nu \beta} k^\mu k^\alpha \right)  \,,
\end{equation}
the tensor longitudinal with respect to the four-momentum $k$
\begin{equation}
  \label{eq:projL}
  P_L = P_1 - P_T  \,,
\end{equation}
the tensor transverse with respect to the four-momentum $k$ and longitudinal with 
respect to the three-momentum\footnote{The three-momentum is measured 
relative to the medium characterized by $p$. A Lorentz covariant expression for
the three-momentum is given by $k-\frac{k\cdot p}{p^2}\,p$.}  $\vec k$
\begin{equation}
  \label{eq:projl}
  P_l^{\mu \nu \alpha \beta} = \frac{1}{2 \, \left(k^2 p^2 - (k\cdot p)^2 \right)} \,
  \left( k^\mu k^\alpha p^\nu p^\beta - k^\mu k^\beta p^\nu p^\alpha 
    - k^\nu k^\alpha p^\mu p^\beta + k^\nu k^\beta p^\mu p^\alpha \right)
\end{equation}
and the tensor transverse with respect to four- and three-momentum
\begin{equation}
  \label{eq:projt}
  P_t = P_T - P_l  \,.
\end{equation}
Obviously all tensors are constructed such that they are antisymmetric with respect
to an exchange of the first (third) and the second (fourth) index. This just reflects
the property of the basic object, the {\em antisymmetric} tensor field $V_{\mu\nu}$.

It is easy to check that the following relations hold:
\begin{subequations}
    \label{eq:projprop}
  \begin{eqnarray}
    P_i \otimes P_i & = & P_i \qquad \mbox{for} \quad i = 1,T,L,t,l \,; \\
    P_1 \otimes P_i & = & P_i \otimes P_1 = P_i \qquad \mbox{for} 
       \quad i = L,T,t,l \,; \\
    P_L \otimes P_i & = & P_i \otimes P_L = 0 \qquad \mbox{for} \quad i = T,t,l \,; \\
    P_T \otimes P_i & = & P_i \otimes P_T = P_i \qquad \mbox{for} \quad i = t,l \,; \\
    P_l \otimes P_t & = & P_t \otimes P_l = 0  
  \end{eqnarray}
\end{subequations}
where the product ``$\otimes$'' is of course defined by contracting the last two 
indices of the
first tensor with the first two indices of the second tensor.

The free propagator is given by \cite{eckgas} 
\begin{equation}
  \label{eq:freeproptensor}
  \langle 0 \vert {\rm T}\, V^{\mu\nu}(x) \, V^{\alpha\beta}(y) \vert 0 \rangle = 
  i \int \frac{d^4\!k}{(2\pi)^4} \, e^{-ik(x-y)} \, \left(
  -\frac{2}{k^2-m^2} \, P_T^{\mu \nu \alpha \beta} 
  + \frac{2}{m^2} \, P_L^{\mu \nu \alpha \beta}
  \right)
\end{equation}
which shows that only transverse (T) modes are propagated while the longitudinal (L)
mode is frozen. Of course, for a free propagator there is no distinction between
transverse (t) and longitudinal (l) polarizations.

Finally we note that we have nothing clever to say about {\em massless} vector mesons:
From \eqref{eq:lagrvec5} it is obvious that the transformations which lead from 
the original Lagrangian \eqref{eq:lagrvec} to \eqref{eq:lagrvec5} only work for
$m \neq 0$. Therefore, the formalism developed in the present section does not work
for massless vector states (and, as already discussed at the beginning of this section, 
also not for the case where the mass is dynamically generated). The formalism does
work, however, for typical hadronic Lagrangians involving massive vector mesons.

\acknowledgments The author acknowledges stimulating discussions with F.~Fr\"omel,
J.~Knoll, A.~Peshier, D.~Rischke and J.~Schaffner-Bielich.
He also thanks U.~Mosel for continuous support.

\appendix

\section{Tadpoles in $\Phi$-derivable approaches}

\begin{figure}[htbp]
    \includegraphics[keepaspectratio,width=0.07\textwidth]{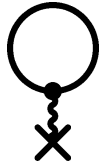}
  \caption{Contribution to the $\Phi$-functional taking into account one-point
functions in the $\phi \varphi^2$ model.
The solid line denotes a propagator of the $\varphi$ mode, the wiggly line plus cross a 
one-point function of the $\phi$ mode. See main text for details.}
  \label{fig:extapp}
\end{figure}

The $\Phi$-derivable scheme deals with full two-point functions. Here tadpoles are 
somewhat special since they are connected to one-point functions. To illustrate
the point we consider a simple model with two distinct scalar fields:
\begin{equation}
  \label{eq:phi3}
{\cal L} = \frac12 \partial_\mu \phi \, \partial^\mu \phi 
+ \frac12 \partial_\mu \varphi \, \partial^\mu \varphi 
- \frac12 M^2 \phi^2 - \frac12 m^2 \varphi^2
- \frac12 g \, \phi \, \varphi^2 \,. 
\end{equation}
In general, the
generating functional to which $\Phi$ contributes must be considered as a functional
of one- and two-point functions and not only of two-point functions. 
It is given by \cite{Ivanov:1998nv}
\begin{equation}
  \label{eq:effact}
\Gamma[\phi_c,\varphi_c,D_\phi,D_\varphi] 
= \underbrace{\int \! d^4x \, {\cal L}[\phi_c,\varphi_c]}_{=S_c[\phi_c,\varphi_c]} 
+ \Phi[\phi_c,\varphi_c,D_\phi,D_\varphi] 
+ \mbox{terms independent of $\phi_c$, $\varphi_c$.}
\end{equation}
Here, ${\cal O}_c$ denote one-point functions (classical fields) and $D_{\cal O}$
propagators. 
$\Phi$ is now given by all two-particle irreducible diagrams. These are all diagrams
which do not fall apart, if two lines are cut. 
\begin{figure}[htbp]
    \includegraphics[keepaspectratio,width=0.3\textwidth]{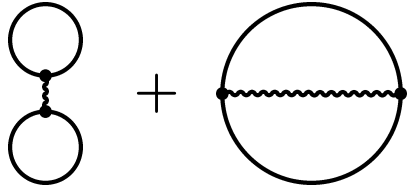}
  \caption{$\Phi$-functional in two-loop order for 
the $\phi \varphi^2$ model. 
Solid lines denote $\varphi$ modes, wiggly lines $\phi$ modes. It will turn out
that the wiggly line in the left diagram must be a bare (non-dynamical) propagator, 
whereas the
wiggly line in the right diagram is a full propagator. See main text for details.}
  \label{fig:Phiapp}
\end{figure}
In particular,
the left diagram of figure \ref{fig:Phiapp} --- the tadpole diagram --- 
does {\em not} enter here, since
it already falls apart after cutting one line. On the other hand, tadpole diagrams do
show up in perturbative evaluations of the self energy. So the question emerges:
Where are the tadpole diagrams in the $\Phi$-approach? Indeed, a tadpole type diagram
does emerge, but it is not the left diagram of figure \ref{fig:Phiapp}. 
Instead, 
the diagram depicted in figure \ref{fig:extapp} must be considered 
(cf.~also \cite{Ivanov:1998nv}). 
Note that
here the wiggly line cannot be cut, since the cross together with the line denotes
one object, namely $\phi_c$. Of course, the right diagram of figure \ref{fig:Phiapp} 
and higher loop orders contribute to $\Phi$. All these diagrams,
however, do not contain $\phi_c$ (in our simple toy model \eqref{eq:phi3}). The
contribution of figure \ref{fig:extapp} to the $\Phi$-functional is given by
\begin{equation}
  \label{eq:contrtadp}
  \Phi_{\rm figure \; \ref{fig:extapp}} = 
  - \frac12 \, \int \! d^4x \, g \, \phi_c(x) \, D_\varphi(x,x)  \,.
\end{equation}

In the following we are aiming at a generating functional for the two-point functions only.
In other words, we want to write down a $\Phi$-functional which only depends 
on the propagators and not any more on the classical fields. Of course, this new
$\Phi$-functional still should yield the
correct self energy for the Dyson-Schwinger equation. To derive this new $\Phi$-functional
all we have to do is to solve the equations of motion for the classical fields and plug
the solutions in $S_c[\phi_c,\varphi_c] +\Phi[\phi_c,\varphi_c,D_\phi,D_\varphi]$ which 
appears in \eqref{eq:effact}. 

The equation of motion for $\varphi_c$ is given by
\begin{equation}
  \label{eq:varphieom}
0 = \frac{\delta \Gamma}{\delta \varphi_c} = - \partial_x^2 \varphi_c(x) - m^2 \varphi_c(x)
- g \phi_c(x) \varphi_c(x)  \,.
\end{equation}
In a thermal system expectation values are independent of the coordinate. We do not
consider spontaneous symmetry breaking for the $\varphi$ mode and conclude: $\varphi_c =0$.
The equation of motion for $\phi_c$ is more involved:
\begin{equation}
  \label{eq:onepointphi}
0 = \frac{\delta \Gamma}{\delta \phi_c} = 
- \partial^2_x \phi_c(x) - M^2 \phi_c(x)
- \frac12 \, g \varphi_c^2(x) - \frac12 \, g D_\varphi(x,x)   \,.
\end{equation}
The last term on the right hand side of (\ref{eq:onepointphi}) 
is exactly the contribution generated from
the diagram in figure \ref{fig:extapp}. Again, in a thermal system also $\phi_c$ is
independent of the coordinate. Using $\varphi_c =0$ one gets
\begin{equation}
  \label{eq:resphic}
\phi_c = - \frac{g}{2M^2} \, D_\varphi(x,x)  \,.
\end{equation}
The factor $1/M^2$ should be interpreted as a {\em bare} (or non-dynamical) $\phi$ 
propagator which couples
to the closed loop $D_\varphi(x,x)$. Since $\phi_c$ can be determined exactly,
one can insert the solution in (\ref{eq:effact}) and obtain a generating functional
for the two-point functions only. The terms caused by $\phi_c$ are
\begin{equation}
  \label{eq:newPhiwoclass}
  S_c[\phi_c,\varphi_c] 
  +\Phi_{\rm figure \; \ref{fig:extapp}}[\phi_c,D_\varphi] 
  = \int \! d^4x \, \left[ - \frac18 \, \frac{g^2}{M^2} \, \left[D_\varphi(x,x)\right]^2 
    + \frac14 \, \frac{g^2}{M^2} \, \left[D_\varphi(x,x)\right]^2 \right]
  = \frac18 \, \int \! d^4x \, \frac{g^2}{M^2} \, \left[D_\varphi(x,x)\right]^2 \,.
\end{equation}
By inspection of this last formula we see that in this way the left diagram of 
figure \ref{fig:Phiapp}
emerges with a subtle, but important aspect: The wiggly line is a {\em bare}
propagator and not a full one.\footnote{This is similar to the non-relativistic case
discussed in \cite{Ivanov:1999tj}. There, in appendix C it is pointed out that also
the classical potential must not be regarded as a dynamical quantity, i.e.~it should not
be considered/cut when calculating derivatives with respect to the full propagator.}
In this way, taking the derivative of this diagram
with respect to $D_\phi$ yields zero, since there is no full $\phi$ propagator.
On the other hand, taking a derivative with respect to $D_\varphi$ yields the
tadpole diagram. 

We have chosen the simple model \eqref{eq:phi3} to illustrate in which way tadpole 
diagrams enter
the $\Phi$-formalism. In our simple case the equations of motion for the classical fields
could be solved exactly. This can be different for more complicated theories. What is
generic, however, is the fact that in a $\Phi$-functional of propagators only, tadpole
diagrams must be included, but with ``tadpole tails'' which are bare and not full
propagators.

The same line of reasoning applies to the models discussed in sections 
\ref{sec:lsm} and \ref{sec:NJL}.

\bibliography{literature,literature2}
\bibliographystyle{apsrev}

\end{document}